\documentclass[11pt]{article}
\usepackage{fleqn,epsf,pslatex,cospar,natbib}

\title{COMPARISON OF EISCAT RADAR DATA ON SPACE DEBRIS WITH MODEL
PREDICTIONS BY THE MASTER MODEL OF ESA}
\author{M. Landgraf\address{ESA/ESOC, Robert-Bosch-Str. 5, 64293
Darmstadt, Germany}, R. Jehn$^{1}$, and W. Flury$^{1}$}
\begin{document}
\maketitle

\begin{picture}(0,0)(0,0)
\put(-50,180){PEDAS1-B1.4-0003-02}
\end{picture}
\begin{abstract}
In the effort to obtain low cost routine space debris observations in
low Earth orbit, ESA plans to utilise the radar facilities of the
European Incoherent Scatter Scientific Association. First
demonstration measurements were performed from 11 to 23 February
2001. In total $16\:{\rm hours}$ of radar signals were collected. Here
we compare these initial measurements with the predictions of the ESA
MASTER/PROOF'99 model in order to assess the sensitivity as well as
the reliability of the data. We find that while the determination of
object size needs to be reviewed, the altitude distribution provides a
good fit to the model prediction. The absolute number of objects
detected in the various altitude bins indicates that the coherent
integration method indeed increases the detection sensitivity when
compared to incoherent integration. In the data presented here
integration times from $0.1$ to $0.3\:{\rm s}$ were used. As expected,
orbit information cannot be obtained from the measurements if they are
linked to ionospheric measurements as planned. In addition routine
space debris observations provide also useful information for the
validation of large-object catalogues.
\end{abstract}

\section*{INTRODUCTION}
Currently ESA monitors the space debris population in low Earth orbit
(LEO) in irregular intervals \citep{mehrholz_thisissue}. This is done
in the frame of 24 hour radar beam-park experiments. The data from
these measurements are taken in order to constrain models of the
population evolution of space debris in orbit \citep{klinkrad01b}. The
most prominent of the European space debris models is MASTER/PROOF'99,
which allows spacecraft operators to assess the risk of a debris
impact for their mission. As the 24 hour experiments only provide a
snapshot of the environment, the temporal behaviour of the space
debris population is difficult to determine. Routine monitoring of LEO
is required in order to better constrain the debris models. The most
suitable technique for monitoring LEO is radar detection by
high-power, large aperture facilities.

In order to obtain inexpensive access to such observations, ESA
started in 2000 a study with the Sodankyl\"a Geophysical Observatory
(SGO) to use the radar facilities of the European Incoherent SCATter
(EISCAT) scientific association. These facilities operate up to 2000
hours per year for ionospheric research purposes. In the study, the
SGO team was able to demonstrate the exploitation of ionospheric
measurements for space debris observations. A comparison of the
retrieved, still preliminary, data with the ESA MASTER/PROOF'99 model
is presented here.

\section*{THE EISCAT FACILITY AT TROMS\O}
The space debris measurements were performed in the frame of a
demonstration campaign in February 2001. A radar facility located at
$69^\circ 35^\prime\:{\rm N}$ and $19^\circ 14^\prime\:{\rm E}$ near
Troms\o, Norway, was used. The UHF radar at Troms{\o} operates at a
frequency of $931\:{\rm MHz}$ with a pulse repetition frequency of
$200\:{\rm kHz}$. The transmitter can transmit peak powers up to
$1.5\:{\rm MW}$ through the main antenna, but it was restricted to
just below $1\:{\rm MW}$ during the measurement campaign. The
$35\:{\rm m}$ antenna was pointed parallel to the field lines of the
local geo-magnetic field, azimuth $183^\circ 18^\prime$ and elevation
$77^\circ 6^\prime$. During the test campaign two different phase
codes called TAU2 and CP1LT were transmitted. Of the TAU2 data $1.2$
hours have been analysed as well as $2.8$ hours of the CP1LT data. For
both experiments the pulse length was about $0.5\:{\rm ms}$. The range
window of the CP1LT experiment covered ranges from $400$ to
$1,400\:{\rm km}$ with a gap between $700$ and $800\:{\rm km}$, and
TAU covered ranges from $500$ to $1,800\:{\rm km}$ with a gap between
$800$ and $1,100\:{\rm km}$.

\section*{CATALOGUED OBJECT}	
\begin{figure}[ht]
\centering
\epsfxsize=.6\hsize
\epsfbox{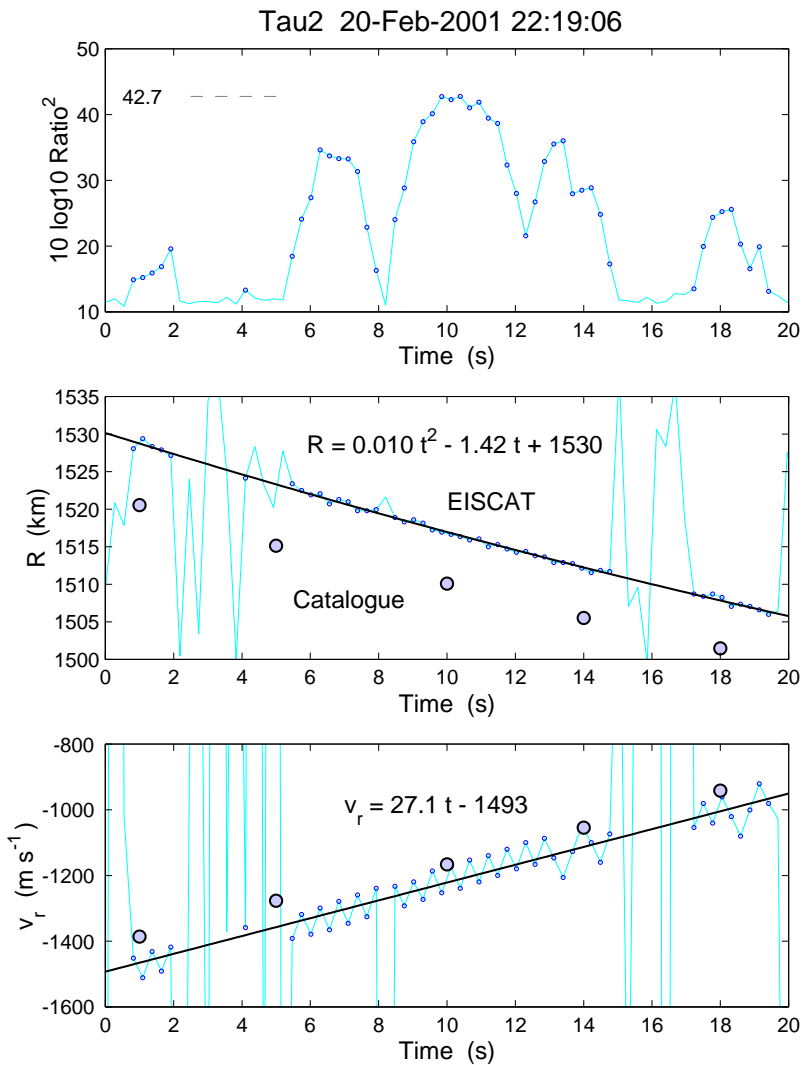}
\caption[Comparison with a catalogued
object]{\label{fig_comparison1}Event at 22:19:06, 20 February 2001,
compared against the catalogue. The top panel shows a quantity
proportional to the signal to noise power ratio. The middle panel
shows the measured slant range (small circles), a parabolic fit, and
the catalogue prediction (large circles). The bottom panel shows the
measured Doppler velocity (small circles), a linear fit, and the
catalogue prediction for the range rate (large circles).}
\end{figure}

The strongest of the high-altitude TAU2 events during the test
campaign began at 22:19:06 UTC, 20~February~2001. We identified the
target as a large catalogued object, a Tsyklon upper stage with the
COSPAR designator 1994-11G. According to the catalogue the object has
a total mass of $1,390\:{\rm kg}$ and is cylindrical in shape, with a
diameter of $2.7\:{\rm m}$, a height of $2.2\:{\rm m}$, and a radar
cross section of $8.3\:{\rm m}^2$. The orbital information in the TLE
catalogue, with TLE epoch one day before the experiment epoch, lets us
predict that it should have passed the radar beam at the off-axis
angular distance of $1^\circ 16^\prime$. Figure~\ref{fig_comparison1}
shows the analysis summary plot of the event. The analysis was done
using $0.27\:{\rm s}$ coherent integration.

The top panel of Figure~\ref{fig_comparison1} shows a quantity
proportional to the signal to noise power ratio during the beam
transit. The antenna side-lobe structure is clearly visible. Markers
indicate those scans where the ratio was larger than the detection
threshold of $4.5$. The off-centre angular distance of the pass was
fit to the data
\citep{EISCAT_FR}. The best fit was achieved by assuming that the
transit occurred $31^\prime$ off-axis. It seems difficult to reconcile
the predicted offset $1^\circ 16^\prime$ with the inferred value of
$31^\prime$.

The middle and bottom panels of Figure \ref{fig_comparison1} show the
measured slant range and Doppler velocity. The solid dark curves
represent quadratic and linear fits to the data points. The large dots
represent the predicted values. The measured range is about $7\:{\rm
km}$ larger than predicted. The slope of the velocity curve is as
predicted, but there is a discrepancy of about $0.1\:{\rm km}\:{\rm
s}^{-1}$ in the actual velocity values. The circular-orbit,
vertical-beam acceleration is $27.5\:{\rm m}\:{\rm s}^{-2}$, which is
consistent with the value $27.1\:{\rm m}\:{\rm s}^{-2}$ from the
velocity fit. The timing accuracy for the catalogued objects is of the
order of $10\:{\rm s}$, while we believe the measured timing to be
accurate to within about $0.1\:{\rm s}$. However, the range and
velocity discrepancies cannot be removed simply by adjusting the
relative timing, because the required correction would be about six
seconds for the range data, but only about three seconds for the
Doppler data. A small misalignment of the antenna can, however,
explain the observed deviation.

This example of the detection of a large catalogued object by the
EISCAT radar demonstrates the usefulness of routine space debris
observations for the validation of large-object catalogues.

\section*{SMALL-SIZE DEBRIS}
While collisions of operational spacecraft with large catalogued
objects like the upper stage discussed above can be avoided by
manoeuvring the spacecraft, objects smaller than $10\:{\rm cm}$, that
normally are not found in the catalogue, pose a serious threat to
manned and unmanned spacecraft. Complex spacecraft like the
International Space Station Alpha are shielded from impacts of objects
smaller than about $1\:{\rm cm}$. Thus, the population in the size
range between $1$ and $10\:{\rm cm}$ is particularly important to
observe.

In the demonstration campaign performed by SGO, $56$ objects in the
size range above $1\:{\rm cm}$ were detected within the $3\:{\rm
hours}$ of analysed data. The measured distribution in size and
altitude distributions discussed below are compared to the prediction
by the ESA MASTER/PROOF'99 model. Due to the pointing of the radar
beam along the local geo-magnetic field line, orbital parameters could
not be determined from the measurements. The predicted distribution of
orbital elements is discussed below anyhow.

\subsection*{Size Distribution}
\begin{figure}[ht]
\centering
\epsfxsize=.5\hsize
\epsfbox{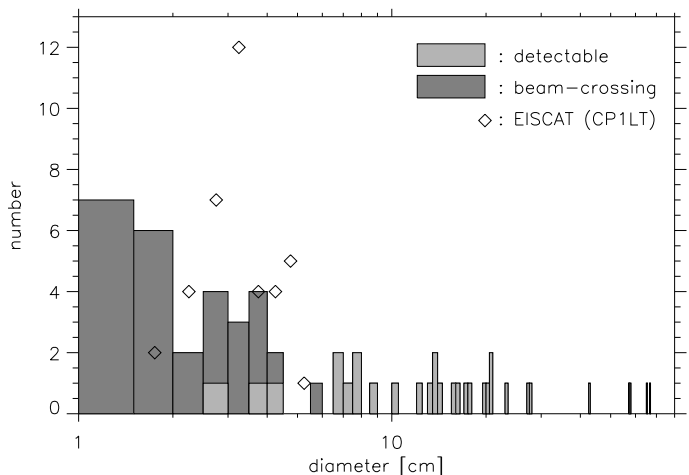}
\caption{\label{fig_sizespec} Distribution of object sizes. The
prediction of the absolute number of beam-crossing objects as a
function of object size by the MASTER/PROOF'99 model is shown as the
dark gray histogram, the light gray histogram shows the detectable
objects, according to the model, and the distribution of sizes of
actually detected objects is shown as the diamond symbols.}
\end{figure}

Figure~\ref{fig_sizespec} shows the distribution of object diameter,
as determined by the analysis of the reflected signal amplitude, as
well as the prediction by the MASTER/PROOF'99 model. Only objects
detected in the range window from $800$ to $1,400\:{\rm km}$ have been
considered.

Under the assumption of incoherent integration, objects larger than
about $7\:{\rm cm}$ are detectable reliably in the range window
between $800$ and $1,400\:{\rm km}$. While the majority of the objects
predicted to cross the beam are smaller than $2\:{\rm cm}$, sizes of
$2$ to $3\:{\rm cm}$ have been derived for the objects actually
detected in the experiment. From the figure it is evident that while
no object with a derived size larger than $6\:{\rm cm}$ has been
detected, more than $20$ of those large objects should have crossed
the beam according to the prediction by MASTER/PROOF'99. Also, the
number of $3\:{\rm cm}$ objects is three times higher than predicted,
while the total number of detected objects is in
agreement. Apparently, the sizes derived for the objects larger than
$10\:{\rm cm}$ have been underestimated. Because it is unknown where
the objects cross the beam pattern, the EISCAT radars can only give a
lower bound for the radar cross section. Consequently, the translation
of the measured radar cross section to object diameter systematically
underestimates the object size.

\subsection*{Altitude Distribution}
\begin{figure}[ht]
\centering
\epsfxsize=.5\hsize
\epsfbox{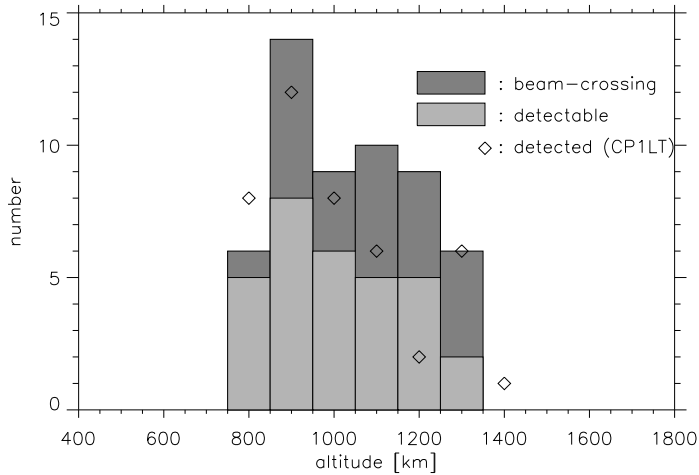}
\caption{\label{fig_altspec} Distribution of orbital altitudes. The
prediction of the absolute object number as a function of orbital
altitude by the MASTER/PROOF'99 model is shown as the gray
histogram. The light gray bars give the number of objects that would
be detectable by the Troms{\o} radar if an incoherent detection method
was used. The dark gray part of each bar indicates the predicted
number of objects undetectable by this method. The distribution of
orbital altitudes of detected space debris object is shown by the
diamond symbols.}
\end{figure}

The distribution of orbital altitudes of detected objects is shown in
Figure~\ref{fig_altspec}. At the altitudes of maximum debris density
between $850$ and $950\:{\rm km}$, $12$ objects were
detected. According to the MASTER/PROOF'99 model, the predicted number
of objects detectable in this altitude bin by incoherent integration
is below $8$, significantly less than actually detected. This is also
true for all but the altitude bin around $1,200\:{\rm km}$. The
sensitivity of the radar appears to be higher than for the standard
incoherent detection method. This can be explained by the fact that
the detection algorithm exploits the phase information by coherently
integrating the detected signal over $0.1\:{\rm s}$. MASTER/PROOF'99
can only simulate radar detection with incoherent integration. Since
the detection algorithm used by the EISCAT team is based on a coherent
integration technique, the predicted number of objects is lower than
the EISCAT data. 

\subsection*{Inclination Distribution}

\begin{figure}[ht]
\centering
\epsfxsize=.6\hsize
\epsfbox{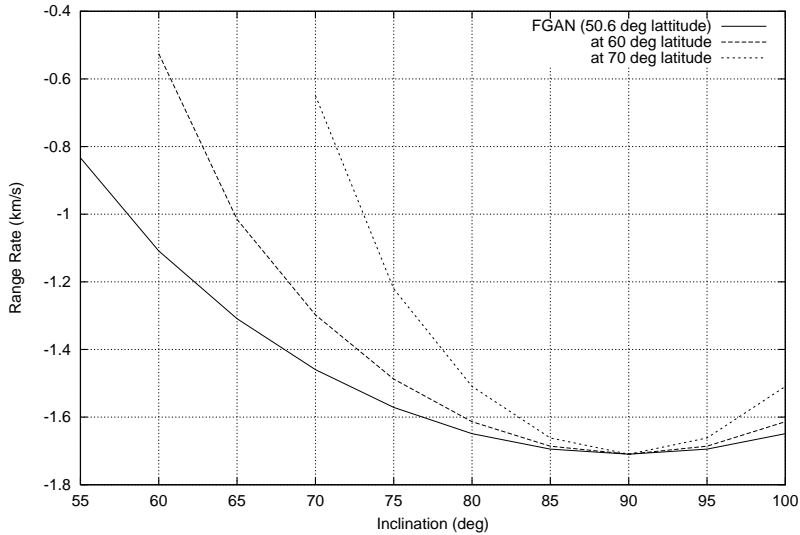}
\caption{\label{fig_rr180} Mapping of orbital inclination to Doppler
velocity for a southern antenna pointing.
}
\end{figure}

\begin{figure}[ht]
\centering
\epsfxsize=.6\hsize
\epsfbox{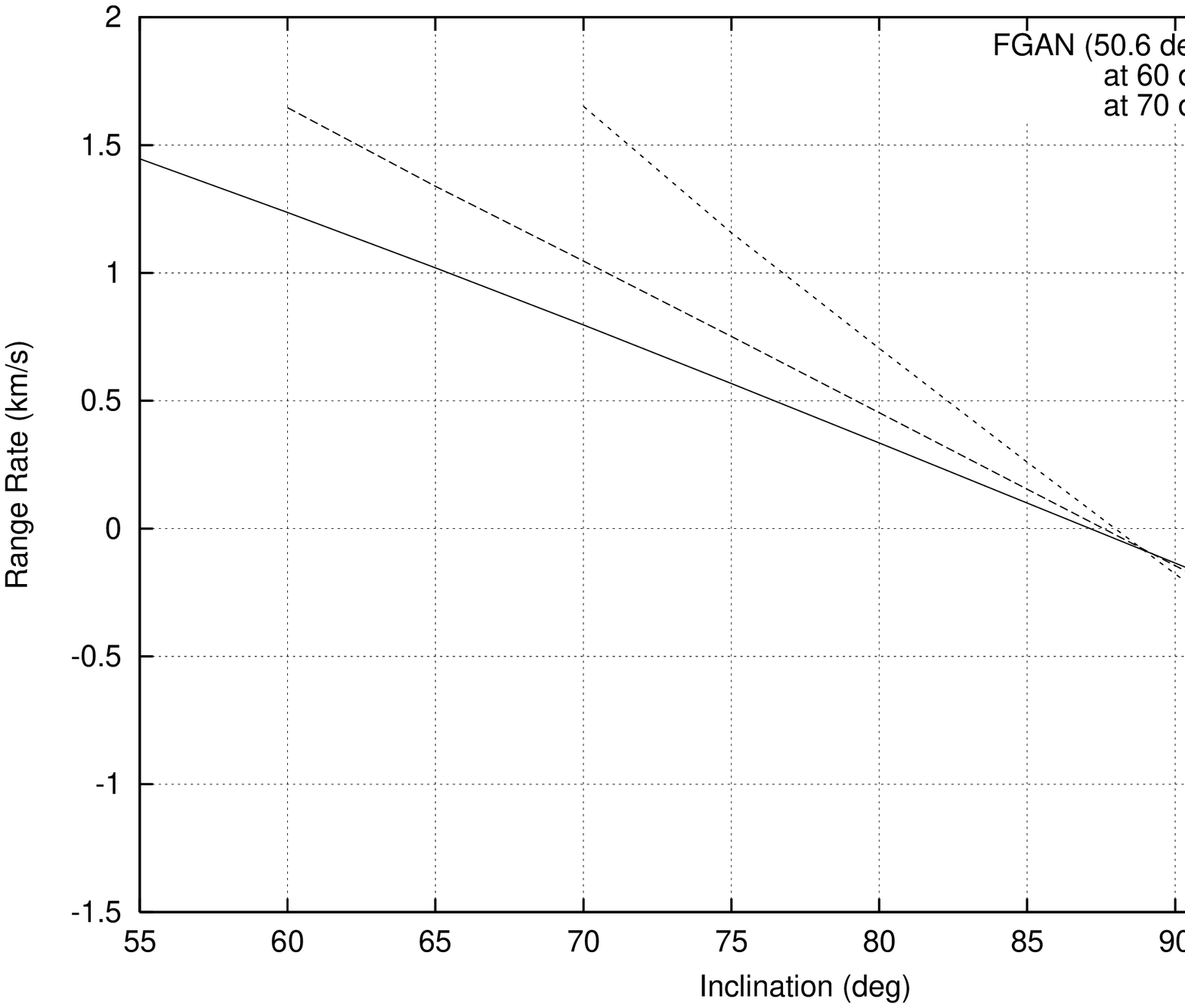}
\caption{\label{fig_rr90} Mapping of orbital inclination to Doppler
velocity for an eastern antenna pointing.
}
\end{figure}

Assuming circular orbits, the inclination of an object's orbit can in
theory be derived from the Doppler velocity measured by the
radar. However, the functional dependence of the Doppler velocity on
the orbit inclination varies with the orientation of the radar
beam. As can be seen in Figures~\ref{fig_rr180} and~\ref{fig_rr90} the
mapping from Doppler velocity to orbit inclination is ambiguous for an
antenna azimuth around $180^\circ$ (South). An easterly pointing
(antenna azimuth $90^\circ$) will provide an unambiguous relationship
between the measured Doppler velocity and orbit inclination.

\begin{figure}[ht]
\centering
\epsfbox{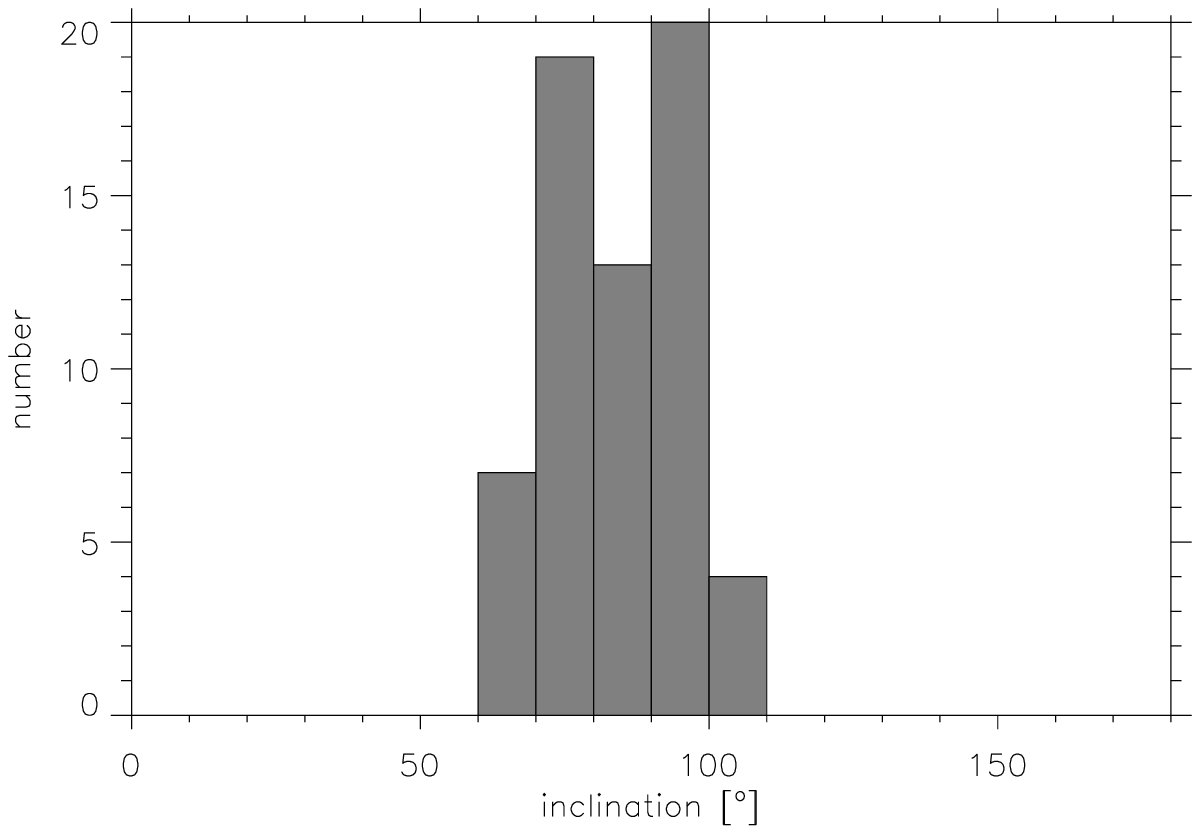}
\caption{\label{fig_incspec} Distribution of orbital inclinations as
predicted by the MASTER/PROOF'99 model.}
\end{figure}
As the EISCAT transmitter at Troms\o\ was pointed parallel to the
geo-magnetic field lines, which follow the north-south direction,
orbit inclinations can not be derived from the data obtained during
the measurement campaign. Using the PROOF application of the MASTER/PROOF'99
model, we can however predict the inclination distribution expected
for the measurement campaign. The prediction is shown in
Figure~\ref{fig_incspec}. It can be seen that inclinations from
$60^\circ$ to $110^\circ$ are covered by beam-park experiments with
the transmitter at Troms\o. This covers debris created in classical
Sun-synchronous orbits as well as debris created by most launches from
Plesetsk.

\section*{CONCLUSION AND OUTLOOK}
The test campaign performed by SGO resulted in raw data on one large
space debris object as well as in the size and altitude distribution
of $55$ small space debris objects. The comparison with the ESA MASTER/PROOF'99
model shows that the conversion from signal amplitudes to object sizes
has to be reviewed. The more straight forward measurement of the
objects' altitude provides a distribution that is well in accord with
the prediction by the model. As the model is regularly checked against
results from other beam-park experiments, we can conclude that the
demonstration measurements with the EISCAT transmitter at Troms\o\
provides reliable measurements. The absolute number of detections is
significantly larger than the predicted number when assuming
incoherent integration as a detection technique. Thus, the coherent
integration of the received signal indeed increases the detection
sensitivity.

As expected, we find that, when linked to ionospheric measurements, we
will not obtain orbital information using the EISCAT radars. This
restriction can, however, easily be removed if dedicated observation
time is acquired. In this case an optimum antenna pointing will allow
us to determine at least Doppler inclinations, which are identical to
true orbit inclinations for circular orbits. As the PROOF tool of the
new MASTER/PROOF-2001 model allows to predict the distribution of Doppler
inclinations, this would allow us to constrain the statistical orbit
properties of the small-size LEO space debris population.

In summary we find that the measurements performed during the
demonstration campaign in February 2001 prove the value of space
debris data obtained by exploiting ionospheric measurements. The mere
amount of data, that will be available in case simultaneous
operations with ionospheric measurements in the order of $1000\:{\rm
hours}$ per year can be performed, provides an important contribution
to the understanding of the evolution of LEO space debris.

In order to realise measurements that operate simultaneously with the
ionospheric investigations, a real-time detection method has to be
developed, because the volume of data that would be needed to record
the raw signal from the radar for $1000\:{\rm hours}$ per year is
prohibitive. With real-time detection only the object parameters as
well as the signal of the passage will be recorded. A study to develop
real-time measurements will be started by the end of 2002.

\end{document}